\documentclass[aps,prc,superscriptaddress,twoside,twocolumn,nofootinbib,showpacs]{revtex4-1}
\usepackage{amsmath,amssymb}
\usepackage{graphicx,bm}
\usepackage{slashed}
\usepackage{dcolumn}
\usepackage{epstopdf}
\usepackage{ulem}
\usepackage[usenames,dvipsnames]{color}
\usepackage{subfigure}
\usepackage{tensor}

\begin{document}

\title{Dynamics of the tri-nuclear system at spontaneous fission of $^{252}$Cf}

\author{R.B. Tashkhodjaev}
\affiliation{National University of Uzbekistan, 100174, Tashkent, Uzbekistan}
\affiliation{Inha University in Tashkent, 100170, Tashkent, Uzbekistan}

\author{A.K. Nasirov}
\email{nasirov@jinr.ru}
\affiliation{National University of Uzbekistan, 100174, Tashkent, Uzbekistan}
\affiliation{BLTP, Joint Institute for Nuclear Research, 141980 Dubna, Russia}

\author{E.Kh. Alpomeshev}
\affiliation{National University of Uzbekistan, 100174, Tashkent, Uzbekistan}

\begin{abstract}
To describe of dynamics of ternary fission of $^{252}$Cf an equation of motion of the tri-nuclear system is calculated. The fission of the $^{70}$Ni+$^{50}$Ca+$^{132}$Sn  channel was chosen as one of the more probable channels of true ternary fission of $^{252}$Cf. The collinearity of ternary fission has been checked by analyzing results of the equation of motion. The results show that if initially all nuclei are placed collinearly (potential energy of this position is the smallest) and the component of the middle fragment's initial velocity which is perpendicular to this line, is zero then ternary fission is collinear, otherwise the non collinear ternary fission takes place.
\end{abstract}


\maketitle

\section{Introduction}

The interest to study of ternary fission was appeared \cite{WJS58} after the discovery of binary fission of heavy nuclei by various authors. Ternary fission is one of the oldest problem of nuclear reaction but still it has topicality. Authors Diehl and Greiner \cite{DG74} tried to explain ternary fission within the framework of the liquid drop model. They mainly calculated the potential energy for the prolate and the oblate configurations of the ternary system. Also, within the three-center shell model Degheidy and Maruhn \cite{Degheidy79} generalized the phenomenological shell model based on the harmonic oscillator potential to systems with three clusters. Authors showed that the centers of nuclei may be in arbitrary geometrical configurations and nuclei may have different masses. The experimental work \cite{Daniel04} was dedicated to study of ternary fission of the $^{252}$Cf nucleus with eight $\bigtriangleup E\times E$ particle telescope. In this experiment, mainly, light charged particles (like He, Be, B etc) were observed in the coincidence with $\gamma$-emission.

The experimental group FOBOS (in the Flerov Laboratory For Nuclear Reactions of the Joint Institute for Nuclear Research, Dubna, Russia) made effort to study the unusual mode of ternary fission - collinear cluster tripartition \cite{PKVA10,PKVA11,PKVA12}. Studies of the spontaneous fission products of $^{252}$Cf in the coincidences with the emitted neutrons has been performed in two missing-mass experiments \cite{PKVA10,PKVA12}. These experiments demonstrated a new mode of the ternary fission process as a collinear cluster tripartition. At first, in collinear cluster tripartition (CCT) mode, the masses of nuclei are comparable, and nuclei are especially clusters, i.e. with a magic number of mass (or charge). At second, the collinearity of the momenta of the ternary fission fragments is proved by the fact that the two detectors registering Sn and Ni-like fragments are placed on the opposite sides from the fissioning source $^{252}$Cf such that angle between them is $180^0$. The probability of the yield of Ni and Sn nuclei are observed was approximately $10^3$ times less then the one of  binary fission. The authors of Ref. \cite{Tash15,NOMT14} concluded that middle fragment is calcium nucleus. This mode of ternary fission differs from usual ternary fission which is binary fission with the emission of light fragments (He, Li, Be etc.) as the third (middle) nucleus in the perpendicular plane to the fission axis.

Moreover some theoretical works dedicated to this kind of ternary fission have been published \cite{TNS11,MB11,VVB12,VBO15,NOMT14,WN14,WNT15,Tash15,NTW16}. In the Ref. \cite{VBO15} the ternary fission of $^{252}$Cf has been studied through the potential energy surfaces for two different arrangements in a collinear configuration, and authors concluded that true ternary fission (with almost equal fragment mass) is energetically possible due to the minima in the fragmentation potential energy and high $Q$-values. Also, in this method it is shown that collinear geometry with the lightest fragment between two heavier nuclei is expected to give the highest probabilities in the decay.

In our previous works \cite{Tash15,WNT15} the possible channels of true ternary fission were studied. In the Ref. \cite{PKVA12} it is shown that more possible channel of ternary fission in the $^{252}$Cf(sf) reaction is $^{70}$Ni+$^{50}$Ca+$^{132}$Sn which is theoretically proven in the Ref. \cite{Tash15}. Our experience from the previous works \cite{Tash15,NTW16} leads to the interesting question: how does tri-nuclear system evaluate during its decay? Because in those works it was not proven that the momenta of fission products are collinear or not. In this work it is decided to study the dynamical changing of relative distance between nuclei and their velocities. So, the main aim of current work is to study of dynamics of fission of the $^{70}$Ni+$^{50}$Ca+$^{132}$Sn system, in other words to check is the ternary fission collinear or not.

Certainly, to get information about dynamics an equation of motion should be solved. Results of solution of the equation of motion depend on initial conditions. The dependence of the result on the initial condition is studied in detail to find the collinear flying of the ternary fission products. Thus from results it will be easy to know what initial condition leads to collinear fission.

\section{The model}

The theoretical model is based on the formation of the tri-nuclear system (TNS). The TNS is the system which has three interacting nuclei \cite{Tash15,NTW16}, and its interaction is studied on the basis of the dinuclear system model \cite{ACNPV93,ACNPV95,NFTA05}. The stage proceeding to formation of the TNS is not studied. It is assumed that the system is formed, and any ternary fission of heavy nuclei passes through the TNS stage.

\begin{figure}
\includegraphics[width=0.6\textwidth]{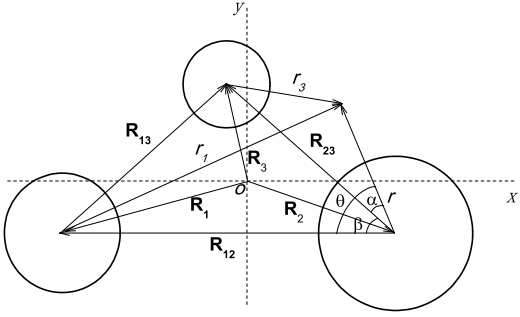}
\vspace*{-3.0cm}
\caption{Point ($\textbf{R}_k$) and relative ($\textbf{R}_{ij}$) vectors of tri-nuclear system. The point $O$ (origin) corresponds to the center of mass.}
\label{graph1}
\end{figure}

The main task is obtaining the classical Lagrange equations of motion, and solving them. First of all, the Lagrangian is $L=T-V$, where $T=\frac{1}{2}\sum^3_{i=1}m_i\dot{\bf{R}}^2_i$ - kinetic energy of the system, $V$ - total interaction potential between fragments. Following system of equations can be written from the Fig. \ref{graph1}:

\begin{equation}
\left\{
\begin{array}{l}
\textbf{R}_{12}=\textbf{R}_1-\textbf{R}_2\\
\textbf{R}_{13}=\textbf{R}_3-\textbf{R}_1\\
\textbf{R}_{23}=\textbf{R}_3-\textbf{R}_2\\
\end{array}\label{Ri-Rij}
\right.
\end{equation}

where $\textbf{R}_k$ ($k=1,2,3$) are point vectors of nuclei and the magnitude of a vector $\textbf{R}_{ij}$ is the relative distance between the i$^{th}$ and j$^{th}$ nuclei.

It is clear that any kind of fission process occurs in one plane, i.e. it can be chosen the 2D space where fission fragments move. So any $\textbf{R}_i$ vector can be described only with $x$ and $y$ components ($R_{ix}$ and $R_{iy}$) in the Cartesian system. Correspondingly, velocities are defined as $\upsilon_{ix}=\dot{R}_{ix}$ and $\upsilon_{iy}=\dot{R}_{iy}$, therefore, the kinetic energy can be written as

\begin{eqnarray}
T=\frac{1}{2}\sum_{i=1}^3m_i(\upsilon^2_{ix}+\upsilon^2_{iy}).
\end{eqnarray}

\subsection{The Lagrange equation of motion}

In the framework of the classical Lagrange formalism 3 equations of motion for the $x$ variable and 3 for the $y$ variable can be obtained:
\begin{eqnarray*}
\frac{d}{dt}\frac{\partial T}{\partial\upsilon_{ix}}-\frac{\partial T}{\partial R_{ix}}=-\frac{\partial V}{\partial R_{ix}}\\
\frac{d}{dt}\frac{\partial T}{\partial\upsilon_{iy}}-\frac{\partial T}{\partial R_{iy}}=-\frac{\partial V}{\partial R_{iy}}
\end{eqnarray*}

It is clear that the kinetic energy does not depend on a distance $R_{i}$, i.e. $\frac{\partial T}{\partial R_{ix}}=\frac{\partial T}{\partial R_{iy}}=0$. Therefore,

\begin{eqnarray}
m_i\dot{\upsilon}_{ix}=-\frac{\partial V}{\partial R_{ix}}\label{EMi}\\
m_i\dot{\upsilon}_{iy}=-\frac{\partial V}{\partial R_{iy}}
\end{eqnarray}

The magnitude of a $\textbf{R}_{ij}$ vector is $R_{ij}=\sqrt{R^2_{ijx}+R^2_{ijy}}$. Potential energy $V$ depends only on relative distance $R_{ij}$ (or $R_{ik}$). So

\begin{eqnarray}
\frac{\partial V}{\partial R_{ix}}=\frac{\partial V}{\partial R_{ijx}}\frac{\partial R_{ijx}}{\partial R_{ix}}+\frac{\partial V}{\partial R_{ikx}}\frac{\partial R_{ikx}}{\partial R_{ix}}=\nonumber\\
=\frac{\partial V}{\partial R_{ij}}\frac{\partial R_{ij}}{\partial R_{ijx}}\frac{\partial R_{ijx}}{\partial R_{ix}}+\frac{\partial V}{\partial R_{ik}}\frac{\partial R_{ik}}{\partial R_{ikx}}\frac{\partial R_{ikx}}{\partial R_{ix}}.
\label{dVdRij}
\end{eqnarray}

It can be noted that $R_{ij}=R_{ji}$ and $R_{ik}=R_{ki}$. If the equation (\ref{EMi}) is written for each nucleus, then using the (\ref{dVdRij}) following equations will be obtained:\\

\begin{equation}
\left\{
\begin{array}{l}
\displaystyle m_1\dot{\upsilon}_{1x}=-\frac{R_{12x}}{R_{12}}\frac{\partial V}{\partial R_{12}}+\frac{R_{13x}}{R_{13}}\frac{\partial V}{\partial R_{13}}\\
\displaystyle m_2\dot{\upsilon}_{2x}=~~\frac{R_{12x}}{R_{12}}\frac{\partial V}{\partial R_{12}}+\frac{R_{23x}}{R_{23}}\frac{\partial V}{\partial R_{23}}\\
\displaystyle m_3\dot{\upsilon}_{3x}=-\frac{R_{23x}}{R_{23}}\frac{\partial V}{\partial R_{23}}-\frac{R_{13x}}{R_{13}}\frac{\partial V}{\partial R_{13}}
\end{array}
\label{FEMx}
\right.
\end{equation}

The relation between $R_{i}$ (or $R_{ix}$) and $R_{ij}$ (or $R_{ijx}$) is found from the system of equations (\ref{Ri-Rij}). Symmetric 3 equations can be gotten for the $y$ component:

\begin{equation}
\left\{
\begin{array}{l}
\displaystyle m_1\dot{\upsilon}_{1y}=-\frac{R_{12y}}{R_{12}}\frac{\partial V}{\partial R_{12}}+\frac{R_{13y}}{R_{13}}\frac{\partial V}{\partial R_{13}}\\
\displaystyle m_2\dot{\upsilon}_{2y}=~~\frac{R_{12y}}{R_{12}}\frac{\partial V}{\partial R_{12}}+\frac{R_{23y}}{R_{23}}\frac{\partial V}{\partial R_{23}}\\
\displaystyle m_3\dot{\upsilon}_{3y}=-\frac{R_{23y}}{R_{23}}\frac{\partial V}{\partial R_{23}}-\frac{R_{13y}}{R_{13}}\frac{\partial V}{\partial R_{13}}
\end{array}
\label{FEMy}
\right.
\end{equation}

Taking into account the conservation law of linear momentum $\displaystyle \sum^3_{i=1}m_i\upsilon_{ix}=0$ (since $\textbf{P}_{c.m.}=0$ for the spontaneous fission of $^{252}$Cf) one of the equations in formulas (\ref{FEMx}) and (\ref{FEMy}) can be skipped. It means that $\upsilon_{3x}$ and $R_{3x}$ is found as

\begin{equation}
\left\{
\begin{array}{l}
\displaystyle \upsilon_{3x}=-\frac{m_1\upsilon_{1x}+m_2\upsilon_{2x}}{m_3}\\
\displaystyle R_{3x}=-\frac{m_1R_{1x}+m_2R_{2x}}{m_3}
\end{array}
\label{V3R3}
\right.
\end{equation}

As the origin is placed at the center of mass so there is no ``$const$'' term in the definition of $R_{3x}$. Equations for $y$ component are similar with the last equation.

\subsection{Derivative of total interaction potential}

It is clear from the Eqs. (\ref{FEMx}), (\ref{FEMy}) and (\ref{V3R3}) that the dynamics of motion strongly depends on derivative of the total interaction potential. The total interaction potential consists of two parts: coulomb and nuclear
\begin{equation}
V=V_C+V_{nuc}.
\label{Vtot}
\end{equation}
As there are three interacting nuclei so there are three terms on each part. To calculate the nuclear part the double folding procedure is used
\begin{equation}
V_C(R_{12},R_{23},R_{13})=e^2\sum^3_{i<j}\frac{Z_iZ_j}{R_{ij}},
\end{equation}
\begin{equation}
V_{nuc}(R_{12},R_{23},R_{13})=\int{\sum^3_{i<j}\rho_i(r_i)f_{ij}(r_i,r_j)\rho_j(r_j)\mathrm{d}\bf{r}}.\label{Vnuc}
\end{equation}
Following formulas are necessary to calculate the nuclear part:
\begin{equation*}
\begin{array}{l}
\displaystyle f_{ij}(r_i,r_j)=C\left[f_{in}+(f_{ex}-f_{in})\frac{\rho_0-(\rho_i+\rho_j)}{\rho_0}\right],\\
\displaystyle \rho_i(r_i)=\frac{\rho_0}{1+\exp[\frac{r_i-R_{0i}}{a}]},\\
\displaystyle r_1(R_{12})=\sqrt{r^2+R_{12}^2-2rR_{12}\cos\theta},\\
\displaystyle r_2=r,\\
\displaystyle r_3(R_{12},R_{23},R_{13})=\sqrt{r^2+R_{23}^2-2rR_{23}\cos\alpha},\\
\displaystyle \cos\alpha=\cos\theta\cos\beta+\sin\theta\sin\beta\sin\phi,\\
\displaystyle \cos\beta=\frac{R_{12}^2+R_{23}^2-R_{13}^2}{2R_{12}R_{23}}.
\end{array}
\end{equation*}
Here, $r_i$ is the radial distance of the i$^{th}$ nucleus (see Fig. \ref{graph1}), $r$, $\theta$ and $\phi$ are variables of the spherical coordinate system, $R_{0i}=r_0A^{1/3}$ - the radius of the i$^{th}$ nucleus, $r_0=1.16$~fm - the radius parameter, $\rho_0=0.17$~fm$^{-3}$ - the density parameter, $a=0.54$~fm - the diffuseness parameter, and $C=300$~MeV$\cdot$fm$^3$, $f_{in}=0.09$, $f_{ex}=-2.59$ are constants of interaction potential, $f_{ij}$ is the effective nuclear-nuclear force which is taken from the Ref. \cite{Migdalbook}.

By the formula (\ref{Vtot}) the derivative of the total interaction potential is found with the two terms
\begin{equation*}
\begin{array}{l}
\displaystyle \frac{\partial V}{\partial R_{ij}}=-e^2\frac{Z_iZ_j}{R^2_{ij}}+\frac{\partial V_{nuc}}{\partial R_{ij}},\\
\displaystyle \frac{\partial V_{nuc}}{\partial R_{ij}}=\int{(F_{12}+F_{23}+F_{13})\mathrm{d}\bf{r}},\\
\displaystyle F_{12}=\rho_2\left[f_{12}-\frac{\rho_1}{\rho_0}C(f_{ex}-f_{in})\right]\frac{\partial \rho_1}{\partial R_{ij}},\\
\displaystyle F_{23}=\rho_2\left[f_{23}-\frac{\rho_3}{\rho_0}C(f_{ex}-f_{in})\right]\frac{\partial \rho_3}{\partial R_{ij}},\\
\displaystyle F_{13}=\rho_3\left[f_{13}-\frac{\rho_1}{\rho_0}C(f_{ex}-f_{in})\right]\frac{\partial \rho_1}{\partial R_{ij}}+\\
\displaystyle +\rho_1\left[f_{13}-\frac{\rho_3}{\rho_0}C(f_{ex}-f_{in})\right]\frac{\partial \rho_3}{\partial R_{ij}},\\
\displaystyle \frac{\partial \rho_1}{\partial R_{ij}}=\frac{\rho_1(\rho_1-\rho_0)}{a\rho_0}\frac{\partial r_1}{\partial R_{ij}},\\
\displaystyle \frac{\partial \rho_3}{\partial R_{ij}}=\frac{\rho_3(\rho_3-\rho_0)}{a\rho_0}\frac{\partial r_3}{\partial R_{ij}}.\\
\end{array}
\end{equation*}
Derivatives $\displaystyle \frac{\partial r_1}{\partial R_{ij}}$ and $\displaystyle \frac{\partial r_3}{\partial R_{ij}}$ are calculated as following
\begin{equation*}
\begin{array}{l}
\displaystyle \frac{\partial r_1}{\partial R_{12}}=\frac{R_{12}-r\cos\theta}{r_1},\\
\displaystyle \frac{\partial r_1}{\partial R_{23}}=\frac{\partial r_1}{\partial R_{13}}=0,\\
\displaystyle \frac{\partial r_3}{\partial R_{12}}=(R_{23}\cos\beta-R_{12})h(r),\\
\displaystyle \frac{\partial r_3}{\partial R_{23}}=\frac{R_{23}-r\cos\alpha}{r_3}-(R_{23}-R_{12}\cos\beta)h(r),\\
\displaystyle \frac{\partial r_3}{\partial R_{13}}=R_{13}h(r),\\
\end{array}
\end{equation*}
where $\displaystyle h(r)=\frac{r}{R_{12}r_3}(\cos\theta-\cot\beta\sin\theta\sin\phi)$.

Be reminded that the integration (\ref{Vnuc}) is provided in the ($x',y',z'$) system, and if $\phi=\pi/2$ then $\theta=\alpha+\beta$ (see Fig. \ref{graph1}).

\section{Results of calculation}

As mentioned above the channel for spontaneous ternary fission of $^{252}$Cf nucleus is chosen as $^{70}$Ni+$^{50}$Ca+$^{132}$Sn. $^{70}$Ni is the first nucleus (placed left side), $^{132}$Sn is the second nucleus (placed right side) and $^{50}$Ca is the third one (placed in the middle) in the Fig. \ref{graph1}. The collinearity of the momenta of the tri-partition is determined by the dynamics of the middle fragment $^{50}$Ca since the heavier fragment $^{132}$Sn is separated firstly and then the middle fragment separates from $^{70}$Ni. This sequence of ternary fission was discussed in the Ref. \cite{Tash15} and it is confirmed by the solution of dynamical equations in this work.

\begin{figure}
\includegraphics[width=0.5\textwidth]{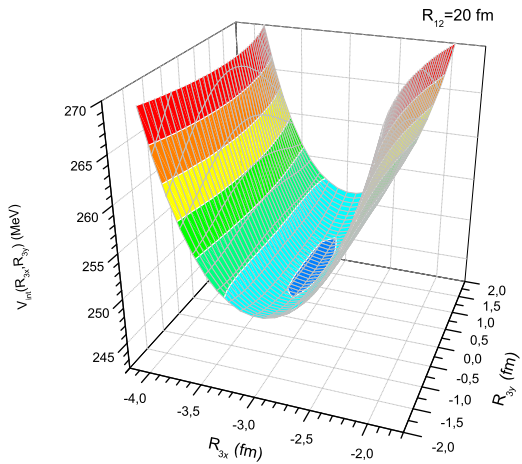}
\vspace*{-0.8cm}
\caption{(Color online). The total interaction potential as the function of $R_{3x}$ and $R_{3y}$ when $R_{12}=20$ fm.}
\label{graph2}
\end{figure}

\begin{figure}
\includegraphics[width=0.5\textwidth]{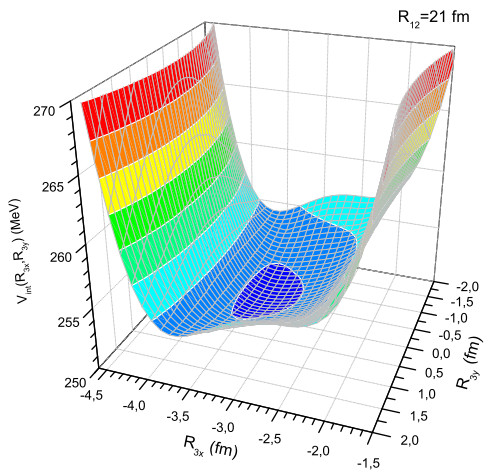}
\vspace*{-0.8cm}
\caption{(Color online). Contour plot of the Fig. \ref{graph2}}
\label{graph3}
\end{figure}

\begin{figure}
\includegraphics[width=0.5\textwidth]{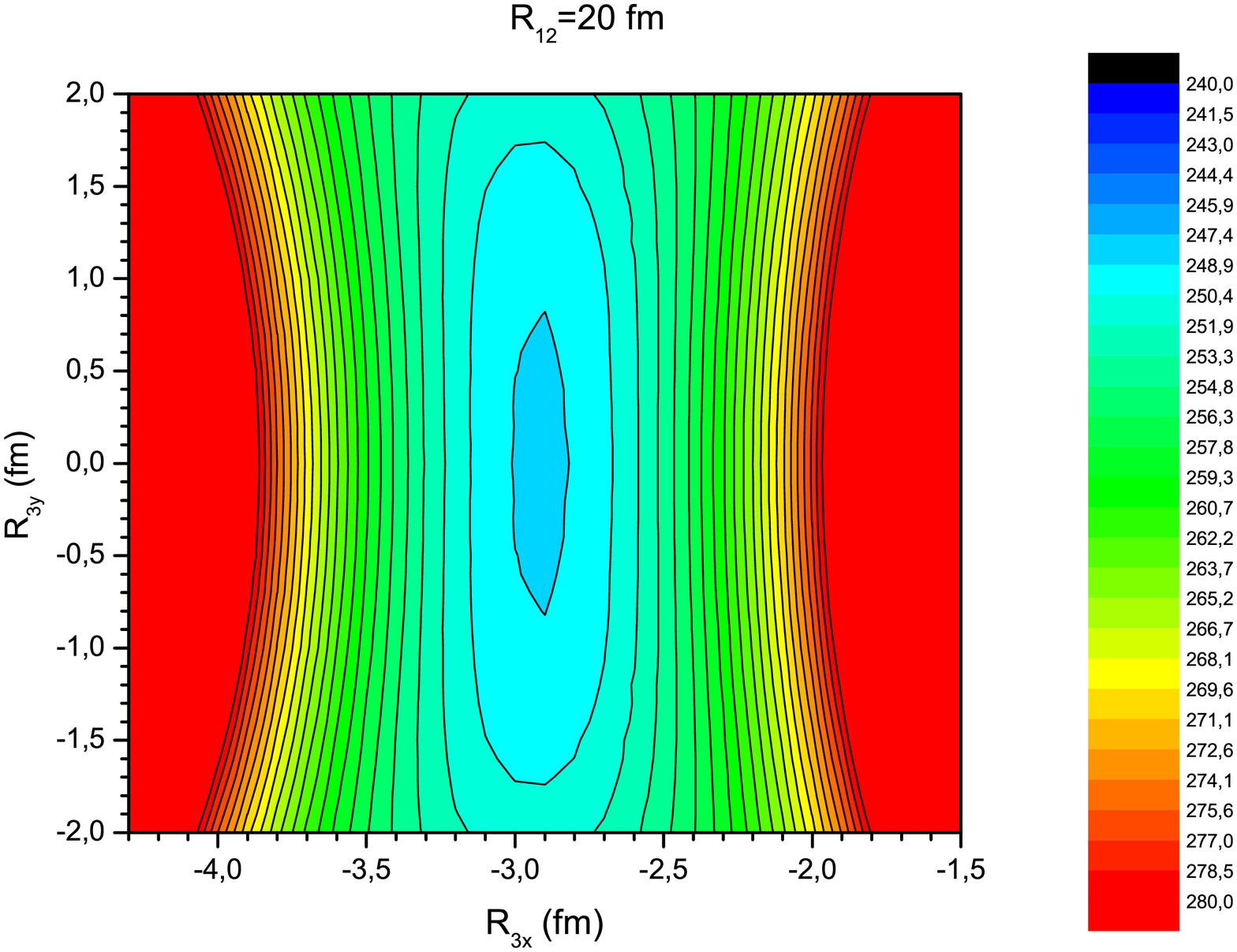}
\vspace*{-0.8cm}
\caption{(Color online). The same as Fig. \ref{graph2} but for $R_{12}=21$ fm.}
\label{graph4}
\end{figure}

\begin{figure}
\includegraphics[width=0.5\textwidth]{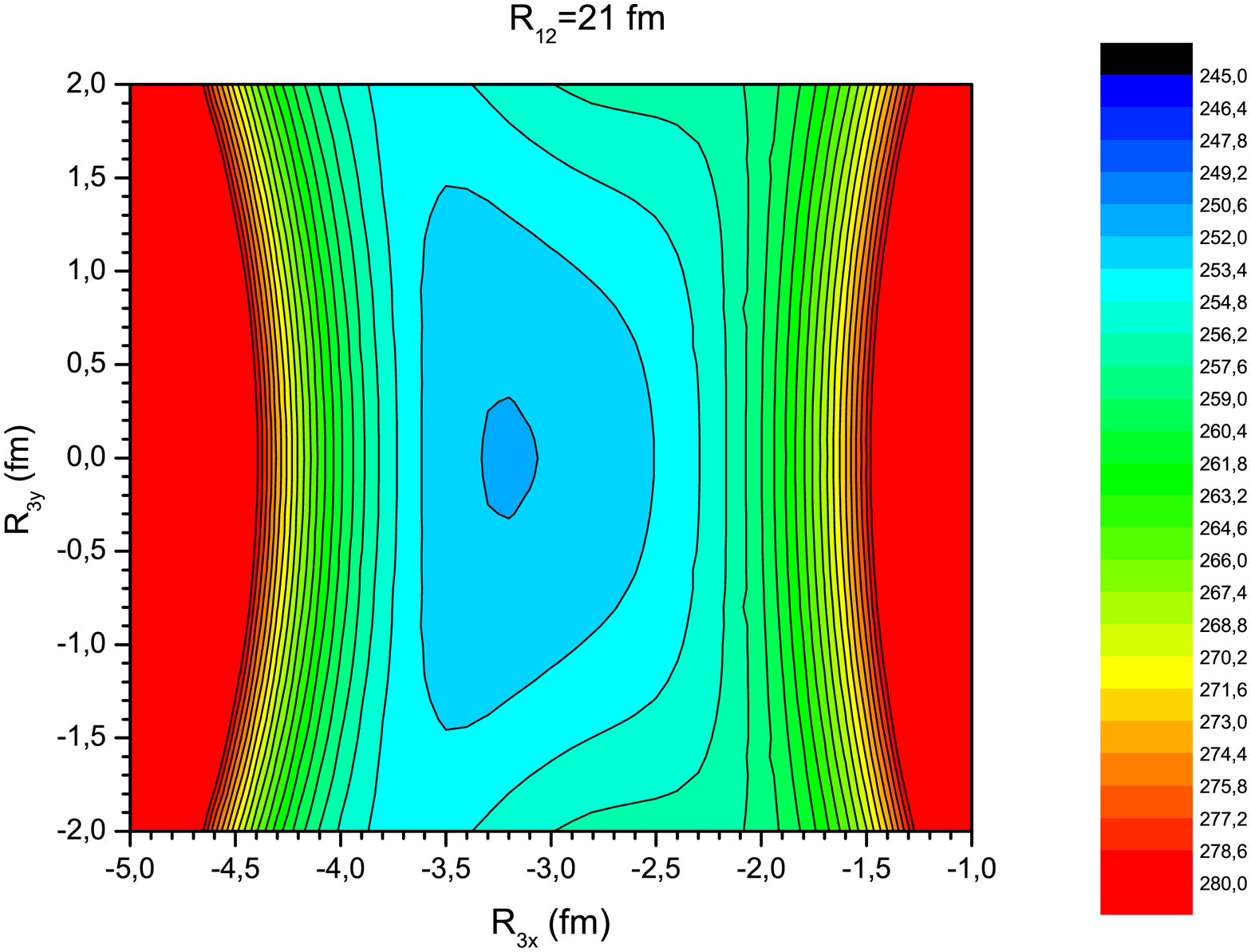}
\vspace*{-0.8cm}
\caption{(Color online). The same as Fig. \ref{graph3} but for $R_{12}=21$ fm.}
\label{graph5}
\end{figure}

\begin{figure}
\includegraphics[width=0.5\textwidth]{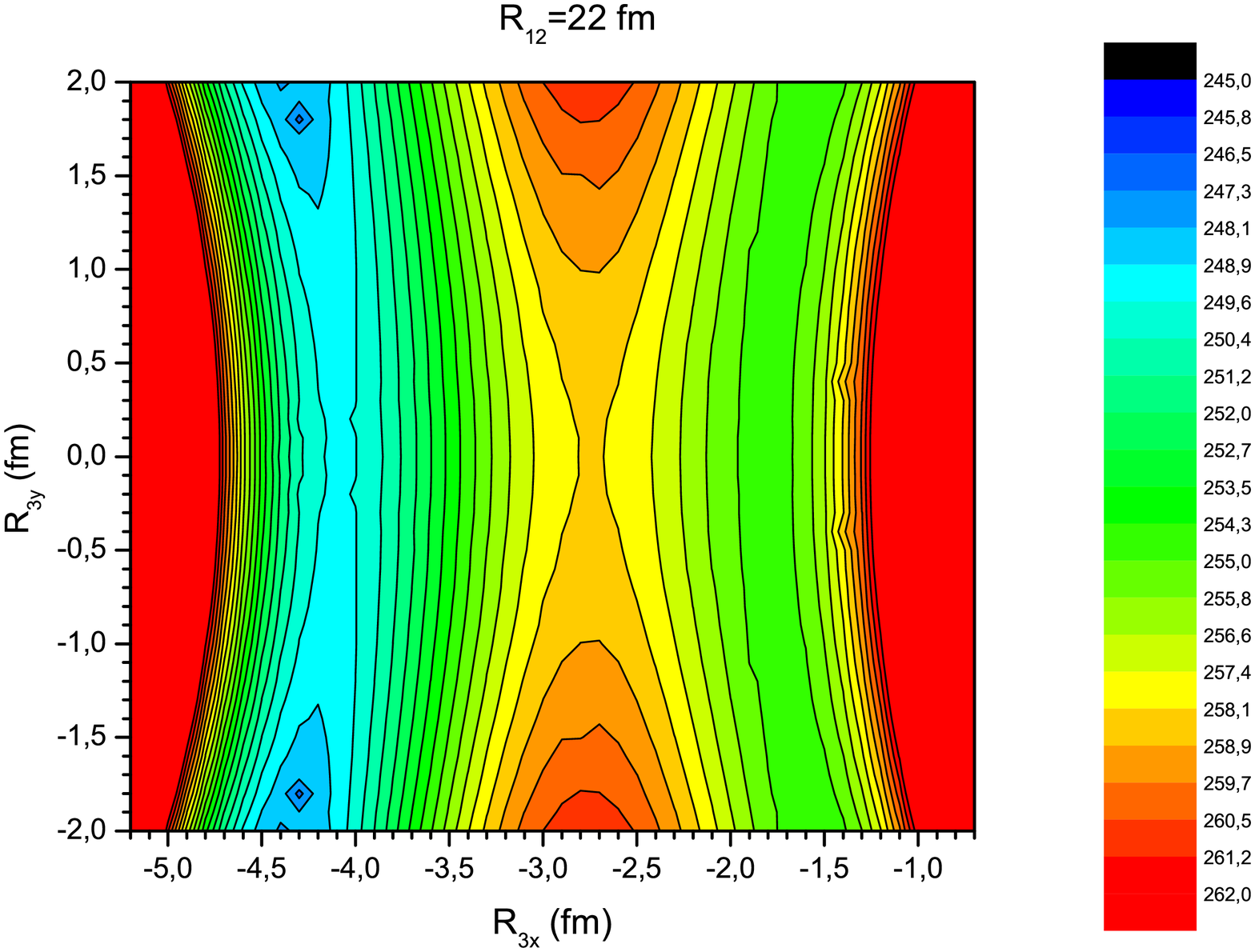}
\vspace*{-0.8cm}
\caption{(Color online). The same as Fig. \ref{graph3} but for $R_{12}=22$ fm.}
\label{graph6}
\end{figure}

\begin{figure}
\includegraphics[width=0.5\textwidth]{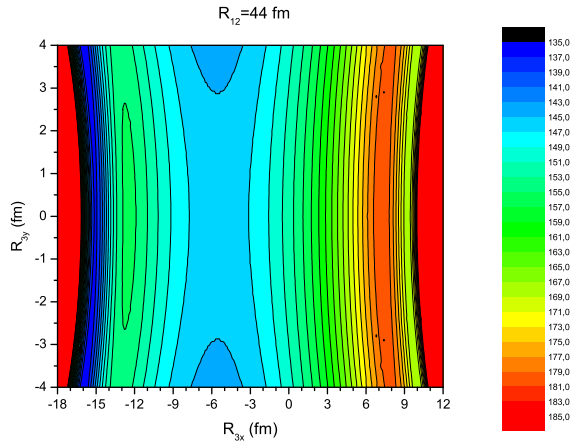}
\vspace*{-0.8cm}
\caption{(Color online). The same as Fig. \ref{graph3} but for $R_{12}=34$ fm.}
\label{graph7}
\end{figure}

It is interesting to discuss how the total interaction potential looks as a function of $R_{3x}$ and $R_{3y}$. It is shown in figures \ref{graph2}-\ref{graph7} for different values of $R_{12}$ (relative distance between Ni and Sn nuclei). The origin (which is not shown) corresponds to the center of mass. There is local minimum at point $R_{3x}=-2.9$ fm and $R_{3y}=0$ fm (see Figs. \ref{graph2} and \ref{graph4}). By increasing of $R_{12}$ the minimum is going to left (to the side of Ni nucleus) but starting from $R_{12}=22$ fm this minimum point is transferred to a saddle point.


\begin{figure}
\includegraphics[width=0.5\textwidth]{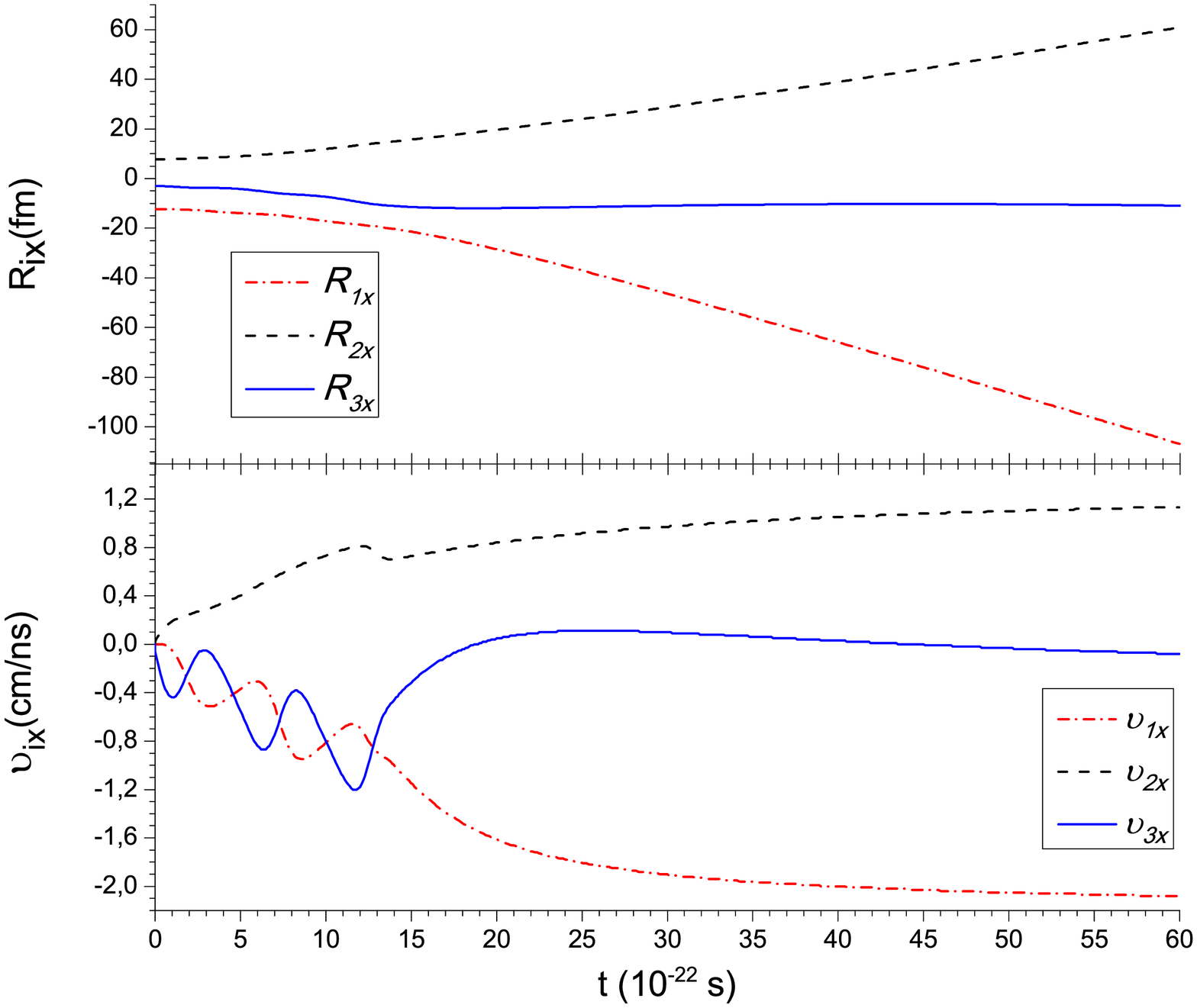}
\vspace*{-0.8cm}
\caption{(Color online). $x$ component of coordinates (upper) and velocities (lower) of three nuclei as functions of time.}
\label{graph8}
\end{figure}

In the first case it is considered that initially all nuclei are placed in one line which means $R_{1y}(t=0)=R_{2y}(t=0)=R_{3y}(t=0)=0$, since the energy of the collinear configuration in the pre-scission state is the smallest, and $x$ coordinates of that nuclei (or relative distance between nuclei) correspond to the local minimum in the Fig. \ref{graph2}, i.e. $R_{1x}(t=0)=-12.3$ fm, $R_{2x}(t=0)=7.7$ fm, $R_{3x}(t=0)=-2.9$ fm. Both components ($x$ and $y$) of initial velocities of the three nuclei are zero. In other words formation of fragments of the TNS so slow that fragments have zero (or too small) velocities. Nevertheless, the assumption of all initial velocities are zero means that there is no the net force which acts to nuclei in the equilibrium state. Results of calculation of the equations of motion (\ref{FEMx}) and (\ref{FEMy}) (together with (\ref{V3R3})) with the mentioned above initial conditions are shown in the Fig. \ref{graph8}. It is shown that from the beginning Sn nucleus is going to breakup from the Ni+Ca system, and then at $t\approx13.5\times10^{-22}~s$ the Ni+Ca system has decayed. Moreover as an important result has been obtained that the third nucleus (Ca) almost does not change its coordinate, because its velocity is about zero. It means the detecting the middle nucleus (Ca) is almost impossible in an experiment. This conclusion proves the assumption done in our previous paper \cite{Tash15}. Only this condition leads to collinear fission of the tri-nuclear system.

\begin{figure}
\includegraphics[width=0.5\textwidth]{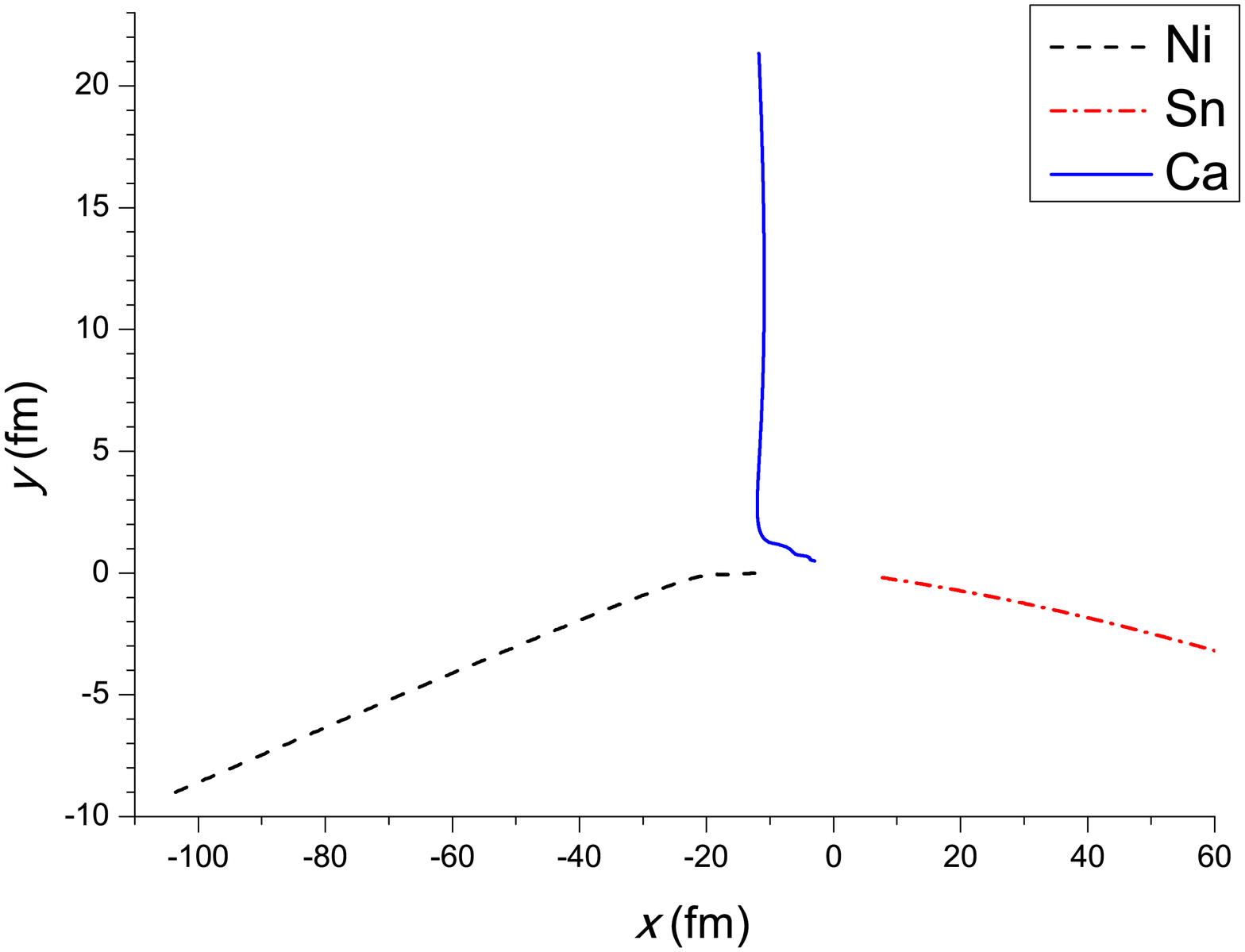}
\vspace*{-0.8cm}
\caption{(Color online). Trajectories of three decaying nuclei when $R_{3y}(t=0)=0.5$ fm and the same initial velocities as in Fig. \ref{graph8}.}
\label{graph9}
\end{figure}

In the second case the velocities of all nuclei are zero, but the middle nucleus (Ca) is placed a little bit upper, i.e. $R_{3y}(t=0)=0.5$ fm, $R_{1x}(t=0)=-12.3$ fm, $R_{2x}(t=0)=7.7$ fm, $R_{3x}(t=0)=-2.9$ fm and $R_{1y}(t=0)=R_{2y}(t=0)=0$. Results of calculation is shown in the Fig. \ref{graph9}. It is clear from the figure that the deviation of the location of calcium nucleus on 0.5 fm on $y$-axis from the origin is enough to get non collinear fission. Moreover, the sequence of the non collinear fission is similar to the one of the collinear fission: at first Sn is separated from Ni+Ca, then, Ni+Ca system is broken up. It is interesting that the decay time of the tri-nuclear system is $t\approx13.2\times10^{-22}~s$ which is almost the same with the time of collinear ternary fission.

\begin{figure}
\includegraphics[width=0.5\textwidth]{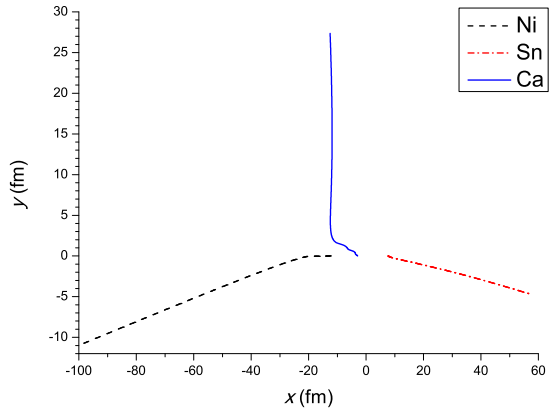}
\vspace*{-0.8cm}
\caption{(Color online). Trajectories of three decaying nuclei when $v_{3y}(t=0)=0.1$ cm/ns and the same initial coordinates as in Fig. \ref{graph8}.}
\label{graph10}
\end{figure}

In the third case initial location of all nuclei are the same as in the first case, i.e. $R_{1y}(t=0)=R_{2y}(t=0)=R_{3y}(t=0)=0$, $R_{1x}(t=0)=-12.3$ fm, $R_{2x}(t=0)=7.7$ fm, $R_{3x}(t=0)=-2.9$ fm. But the initial velocity of the middle fragment is $v_{3y}(t=0)=0.1$~cm/ns, and other initial velocities are zero: $v_{1x}(t=0)=v_{2x}(t=0)=v_{3x}(t=0)=v_{1y}(t=0)=v_{2y}(t=0)=0$. The Fig. \ref{graph10} shows that if $y$ component of the initial velocity of the Ca nucleus is not zero, ternary fission will be non collinear. Also, it should be noted that in this case the decay time of the TNS is $t\approx13.4\times10^{-22}~s$ which is almost the same with the time of the previous cases. From the comparison of Figs. \ref{graph9} and \ref{graph10} is seen that the final paths of the all fragments are similar in spite of difference in the initial conditions of the second and third cases.

From the figures it can be concluded that there is collinear fission only when all three nuclei are located in one line ($R_{iy}=0$) and there is not $y$ component of the initial velocity of the middle fragment ($v_{3y}=0$).

\section{Summary}

We conclude that if in the pre-scission stage all nuclei are placed collinearly which corresponds to the minimum in the potential energy surface and there is no the net force on the third nucleus (Ca) on y-axis (or $y$ component of its initial velocity is zero), then the tri-nuclear system can be broken up collinearly. This theoretical result proves the experimental results of the collinear cluster tripartition in the Ref. \cite{PKVA12}. The experiment shows that collinear ternary fission can be observed. Therefore, in the framework of the TNS model the initial condition which leads to collinear fission have a place in the nature.

From the comparison of the potential energy surfaces in the Figs. \ref{graph2}-\ref{graph7} it can be concluded that as $R_{12}$ (relative distance between Ni and Sn nuclei) increases the minimum at the point when $R_{3x}=-2.9$ fm and $R_{3y}=0$ fm in the Fig. \ref{graph2} is disappeared, and instead of this minimum the saddle point is emerged (see Fig. \ref{graph7}). It means the TNS with value of $R_{12}$ higher then 22 fm is an unstable system.

Moreover, from the Figs. \ref{graph9} and \ref{graph10} as conclusion it can be emphasized that non collinear ternary fission occurs in the following initial conditions: the deviation in y-axis of location from the origin of the middle (Ca) nucleus or the difference from zero of the $y$ component of the velocity of that nucleus.

Nevertheless, it is interesting that in all cases the decay time of the TNS has nearly the same value. It means the time almost does not depend on initial conditions. It is because of the sequence of the fission: firstly, Sn nucleus is separated from the Ni+Ca system, and then Ni is decayed from Ca nucleus.

As the collinearity of the ternary fission depends on initial conditions so the probability (or the weight) of the each initial condition's population is the open question which will be studied in future investigations.

\acknowledgments

This paper has done in the framework of the Project F2-FA-F115 of the Committee for coordination science and technology development under the Cabinet of Ministers of Uzbekistan. The authors A.K.N. and R.B.T. thank to the committee for their financial support. A.K.N. thanks the Russian Foundation for Basic Research for the partial support.

\end{document}